\documentclass{aa}
\usepackage{graphicx}
\usepackage{txfonts}
\newcommand{\RSTAR}{\mbox{$R_{\star}$}}
\newcommand{\TEFF}{\mbox{$T_{\rm eff}$}}
\newcommand{\TIN}{\mbox{$T_{\rm in}$}}
\newcommand{\LOGG}{\mbox{$\log \varg$}}

\newcommand{\RSOL}{\mbox{$R_{\sun}$}}
\newcommand{\MSOL}{\mbox{$M_{\sun}$}}
\newcommand{\LSOL}{\mbox{$L_{\sun}$}}
\newcommand{\MSOLPERYR}{\mbox{$M_{\sun}$~yr$^{-1}$}}
\newcommand{\micron}{\mbox{$\mu$m}}
\newcommand{\KMS}{\mbox{km s$^{-1}$}}
\newcommand{\RIN}{\mbox{$r_{\rm in}$}}
\newcommand{\RTRANS}{\mbox{$r_{\rm tr}$}}
\newcommand{\ROUT}{\mbox{$r_{\rm out}$}}
\newcommand{\TAUV}{\mbox{$\tau_{V}$}}
\newcommand{\fgsge}{\mbox{FG~Sge}}
\begin{document}

\title{
  Spatially resolved mid-infrared observations of the circumstellar
  environment of the born-again object FG~Sge
\thanks{
Based on observations made with the Very Large Telescope Interferometer of 
the European Southern Observatory. Program ID: 081.D-0244}
}

\author{K.~Ohnaka\inst{1}
\and
B.~A.~Jara Bravo\inst{1}
}

\offprints{K.~Ohnaka}

\institute{
  Instituto de Astrof\'isica, Universidad Andr\'es Bello,
  Fern\'andez Concha 700, Las Condes, Santiago, Chile\\
\email{k1.ohnaka@gmail.com}
}

\date{Received / Accepted }

\abstract
{
FG~Sge has evolved from the hot central star of the young planetary nebula
Hen 1-5 to a G--K supergiant in the last 100 years.  It is one of the three 
born-again objects that has been identified as of yet, and they are considered to 
have undergone a thermal pulse in the post-asymptotic giant branch 
evolution. 
}
{
We present mid-infrared spectro-interferometric observations of FG~Sge and 
probe its dusty environment.
}
{
  FG~Sge was observed with MIDI at the Very Large Telescope Interferometer
  at baselines of 43 and 46~m between 8 and 13~\micron. 
}
{
  The circumstellar dust environment of FG~Sge was spatially resolved, and
  the Gaussian fit to the observed visibilities results in a
  full width at half maximum of $\sim$10.5~mas. 
  The observed mid-infrared visibilities and the 
  spectral energy distribution can be fairly reproduced by optically thick
  (\TAUV\ $\approx 8$) spherical dust shell models consisting of amorphous
  carbon with an inner radius \RIN\ of $\sim$30~\RSTAR\ (corresponding to
  a dust temperature of $1100\pm100$~K). 
  The dust shell is characterized with a steep density profile proportional
  to $r^{-3.5\pm0.5}$ from the inner radius \RIN\ to $(5-10)\times\RIN$, 
  beyond which it changes to $r^{-2}$. 
  The dust mass is estimated to be $\sim \! 7 \times 10^{-7}$~\MSOL,
  which translates into an average total mass-loss rate of
  $\sim \! 9 \times 10^{-6}$~\MSOLPERYR\ as of 2008 with a gas-to-dust
  ratio of 200 being adopted.
  In addition, 
  the 8--13~\micron\ spectrum obtained with MIDI with a field of view of
  200~mas does not show a signature
  of the polycyclic aromatic hydrocarbon (PAH) emission,
  which is in marked contrast to the
  spectra taken with the Spitzer Space Telescope six and 20 months before
  the MIDI observations with wide slit widths of 3\farcs6--10\arcsec.
  This implies that the PAH emission originates from an extended
  region of the optically thick dust envelope.
}
{
  The dust envelope of \fgsge\ is much more compact than that of the other
  born-again stars'
  Sakurai's object and V605~Aql, which might reflect
  the difference in the evolutionary status. 
  The PAH emission from the extended region of
  the optically thick dust envelope likely originates from the material
  ejected before the central star became H-deficient, and 
  it may be excited by the UV radiation
  from the central star escaping through gaps among dust clumps and/or
  the bipolar cavity of a disk-like structure. 
}


\titlerunning{Spatially resolved mid-infrared observations of FG~Sge}
\authorrunning{Ohnaka \& Jara Bravo}
\maketitle

\section{Introduction}
\label{sect_intro}

FG~Sge is the central star of the young planetary nebula Hen~\mbox{1-5} and 
a very rare object whose evolution can be observed in a lifetime 
of a human being.  
In the last 100 years, it has rapidly evolved redward on the 
Hertzsprung-Russell (H-R) diagram,
from the hot central star of the planetary nebula (effective
temperature $T_{\rm eff} \approx 40\,000$~K) to a G supergiant 
($T_{\rm eff} \approx 5500$~K, Jeffery \& Sch\"onberner \cite{jeffery06}). 
The rapid evolution of FG~Sge is considered to be caused by a late thermal 
pulse (LTP) that occurs after the asymptotic giant branch (AGB),
when the star is 
evolving blueward and almost horizontally on the H-R diagram.
The nuclear energy produced by the thermal pulse results in 
the expansion of the thin outer layers on a timescale of 50 to a few 100 
years, and the star evolves rapidly redward on the H-R diagram, turning into 
a so-called born-again object (e.g., Langer et al. \cite{langer74}; 
Iben \cite{iben76}; Bl\"ocker \cite{bloecker01}).  
When the temperature of the expanding outer layers becomes 
low enough for convection to penetrate to deep layers, the nuclear-processed 
material, such as C, He, and {\sl s}-process elements, is dredged up to 
the surface.  Since the mass of the H-rich outer layers is already 
small, this dredge-up process makes the atmosphere H-deficient and C-rich.  
After the surface temperature reaches its minimum of several 1000~K, 
the star evolves again 
blueward on a timescale of $10^3$--$10^4$ years finally to a white dwarf.  

The final flash episode is so brief that objects such as
FG~Sge are very rare.
Only three objects are known to date: 
FG~Sge, Sakurai's object (V4334~Sgr), and V605~Aql. 
The evolution of \fgsge\ back to the AGB over $\sim$100 years is
consistent with the LTP scenario, 
while the return of Sakurai's object and V605~Aql to the AGB is even more
rapid on the order of several years. 
They are considered to have experienced
the so-called very late thermal pulse (VLTP) -- a thermal pulse that
occurs when the star is evolving downward to a white dwarf on the H-R
diagram (Clayton et al. \cite{clayton06}). 
In addition, Reindl et al. (\cite{reindl17}) recently found that SAO244567,
the central star of the Stingray nebula, has been rapidly returning to
the AGB since 2002, and they conclude that it has undergone an LTP. 
The final flash episode is considered to be a fairly common 
phenomenon, which occurs in 10--20\% of stars evolving off the AGB 
(Bl\"ocker \cite{bloecker01}; Herwig \cite{herwig05}).  
Therefore, studies of final flash objects provide a unique opportunity 
to obtain insight into the evolution of post-AGB stars.  

Although \fgsge\ is discussed as an LTP object in the literature
(e.g., Sch\"onberner \cite{schoenberner08}), the evolution of its chemical
composition is not yet fully understood.
The detailed spectroscopic analysis of Jeffery \& Sch\"onberner 
(\cite{jeffery06}) shows that the atmosphere of FG~Sge became 
H-deficient some time between 1960 and 1995, and the detection 
of C$_2$ (Iijima \& Strafella \cite{iijima93}; 
Kipper \& Kipper \cite{kipper93}) reveals that the atmosphere 
is C-rich.  
However, this photospheric chemical composition of FG~Sge and its evolution on 
the H-R diagram cannot be consistently explained by the present stellar 
evolutionary theory as discussed by Jeffery \& Sch\"onberner
(\cite{jeffery06}).
The calculations of born-again 
giants (Bl\"ocker 2001) predict that the atmosphere becomes 
H-deficient only
    {after} the star begins to evolve blueward again 
(i.e., when the temperature is increasing again). 
However, FG~Sge became H-deficient between 1960 and 1995, when the 
surface temperature was still decreasing.  
In the case of a VLTP, 
hydrogen in the outer layers becomes significantly depleted 
due to the hydrogen burning {during} the helium flash
(Herwig et al. \cite{herwig99}). 
This means that the star is already H-deficient when it begins to 
evolve redward.  However, FG~Sge was H-normal in 1960 when it was 
evolving redward.  
The alternative double-loop scenarios proposed by
Lawlor \& MacDonald (\cite{lawlor03})
and Miller Bertolami et al. (\cite{millerbertolami06}), in which 
\fgsge, Sakurai's object, and V605~Aql can be considered to 
represent different evolutionary
stages of the same evolutionary path,
cannot entirely explain the observed chemical composition of \fgsge\ either,
as Jeffery \&  Sch\"onberner (\cite{jeffery06}) have discussed.

Another remarkable change occurred in 
FG~Sge in 1992,
when its visual brightness decreased suddenly by
$\Delta V \approx 5$~magnitude 
and recovered gradually over the course of $\sim$1 year.  Since 
then, FG~Sge has shown sudden, unpredictable deep declines at an 
interval of 1--2~years; although, such events have become so frequent that the
central star has been seldom visible since 2008 
(Arkhipova et al. \cite{arkhipova22}). 
At the same time, it started to show a noticeable 
infrared excess characterized by carbonaceous dust with $\sim$1000~K 
(Gehrz et al. \cite{gehrz05}).
The light curve of FG~Sge is similar to those of R~CrB stars -- 
a class of F--G supergiants
in which optically thick dust clouds are ejected in random directions -- and 
an occasional, deep decline is observed when a cloud happens to form in the 
line of sight (Clayton \cite{clayton96}).  
While some authors propose that FG~Sge is undergoing such a random 
dust cloud formation (Gonzalez et al. \cite{gonzalez98}), 
Gehrz et al. (\cite{gehrz05}) have argued for a continuous outflow with abrupt,
episodic changes of the mass-loss rate. 
In either case, however, it is unknown where and how dust can form 
around FG~Sge, whose present temperature (4500--5500~K) is deemed to be too 
high for dust to condensate.  

To obtain a clearer picture of the dusty environment of FG~Sge,
high spatial resolution mid-infrared observations are effective.  
In this paper, we present the first spatially resolved mid-infrared
(8--13~\micron) observations of \fgsge\ and radiative transfer modeling of
its circumstellar envelope.

\section{Observations}
\label{sect_obs}

FG~Sge was observed with the MIDI instrument (Leinert et al. \cite{leinert03})
at ESO's Very Large Telescope Interferometer (VLTI) 
on 2008 June 22 (MJD54639) using the
unit telescope (UT)
configuration UT2--UT3 with projected baseline lengths of 43--46~m 
(Program ID: 081-D0224, P.I.: K.~Ohnaka). 
A prism with a spectral resolution of $\lambda/\Delta \lambda \simeq 30$ 
(at 10~\mbox{$\mu$m}) was used to obtain spectrally dispersed fringes
in the $N$ band between 8 and 13~\mbox{$\mu$m}. 
The data were taken in SCI-PHOT mode, in which photometric data of each
telescope are recorded simultaneously with interferometric data. 
A detailed description of the observing procedure is given in 
Przygodda et al. (\cite{przygodda03}), Leinert et al.\ (\cite{leinert04}), 
and Chesneau et al. (\cite{chesneau05}).  
Table~\ref{table_obs} summarizes the observations' \fgsge\ and the
calibrator. 

\begin{table}
\begin{center}
\caption {MIDI observations of \fgsge\ and the calibrator: 
time of observation (coordinated universal time=UTC), projected baseline
length $B_{\rm p}$, position angle 
of the projected baseline on the sky (P.A.), seeing in the visible, and the
airmass at the times of the observations of \fgsge\ and the calibrator.
}
\begin{tabular}{l l c c r r l}
\hline
\hline
\# & Object & $t_{\rm obs}$ & $B_{\rm p}$ & P.A.   & Seeing & Airmass\\ 
   &        & (UTC)       & (m)       & (\degr) & (\arcsec) & \\
\hline
\multicolumn{6}{c}{2008 June 22 (MJD2454639)} \\
\hline
CAL1  & HD168723 & 07:27:42 & 46.0 & 46  & 1.0  & 1.35 \\
SCI1  & \fgsge   & 07:46:02 & 43.0 & 50  & 1.0  & 1.47  \\
CAL2  & HD168723 & 08:36:15 & 43.3 & 43  & 1.0  & 1.84  \\
SCI2  & \fgsge   & 08:57:21 & 45.9 & 47  & 1.0  & 1.74  \\
SCI3  & \fgsge   & 09:06:47 & 46.1 & 46  & 1.1  & 1.82  \\
\hline
\label{table_obs}
\end{tabular}
\end{center}
\end{table}

We used the MIA+EWS package ver.2.0\footnote{Available at 
http://www.strw.leidenuniv.nl/\textasciitilde nevec/MIDI} 
to reduce the MIDI data (Leinert et al. \cite{leinert04}; 
Jaffe \cite{jaffe04}).  
We adopted a uniform-disk diameter of 3.1~mas for the calibrator
HD168723 (Bourges et al. \cite{bourges17}). 
It should be noted that there are significant differences in airmass
among the data sets.
To avoid systematic
errors, we calibrated the data sets of \fgsge\ with those of the calibrator
taken at a similar airmass: the data set SCI1 was calibrated with CAL1, while
the data sets SCI2 and SCI3 were calibrated with CAL2. 

We assessed the data quality and derived the calibrated visibilities as 
described by Ohnaka et al. (\cite{ohnaka08}).  
Because it is difficult to estimate the errors in the calibrated 
visibilities properly from two calibrator data sets taken at very different
airmasses, we assumed a total relative error of 10\% as adopted 
by Zhao-Geisler et al. (\cite{zhao-geisler11}) for SCI-PHOT mode data. 
Because the projected baseline length and position angle of 
SCI2 and SCI3 are very close, we averaged the calibrated visibilities
of these two data sets. 
The visibilities obtained with MIA and EWS agree well, and 
we only provide the results derived with EWS in the discussion 
below.  

The absolutely calibrated $N$-band spectra of FG~Sge were obtained 
from the same three SCI--CAL pairs taken at similar airmasses 
(SCI1--CAL1, SCI2--CAL2, and SCI3--CAL2), using the method 
described in Ohnaka et al. (\cite{ohnaka07}).
The average and the standard deviation of the three spectra were taken as
the final spectrum and its error, respectively.

We also obtained the 19.5~\micron\ flux of FG~Sge with the mid-infrared
imager and spectrograph VLT/VISIR (Lagage et al. \cite{lagage04}).
The VISIR observations of FG~Sge and a 
photometric standard star, HD173780, were carried out on 2008 May 30, 
about three weeks before the MIDI observations, with the Q3 filter
(central wavelength = 19.5~\micron, filter full width at half maximum 
= 0.4~\micron) and a pixel scale of 0\farcs075. 
The data with chopping and nodding were first reduced using the VISIR
pipeline ver.3.0.0 to obtain the images with the sky background almost
subtracted. From these images, we further subtracted the residual sky
background measured in an annular region with an inner and outer radius
of 1\farcs5 and 2\farcs25, respectively. 
The flux was obtained within an aperture radius of 1\farcs5 both for
\fgsge\ and HD173780, and the data of \fgsge\ were flux-calibrated with
a Q3-band flux of 1.7276~Jy for 
HD173780\footnote{https://www.eso.org/sci/facilities/paranal/instruments/visir/tools/\\
  zerop\_cohen\_Jy.txt}. 
The error in the calibrated flux of \fgsge\ was estimated by the quadratic
sum of the photon noise of the source and that of the background. 
The observed 19.5~\micron\ flux of \fgsge\ is $6.2 \pm 0.49$~Jy.

\begin{figure}[!t]
\resizebox{\hsize}{!}{\rotatebox{0}{\includegraphics{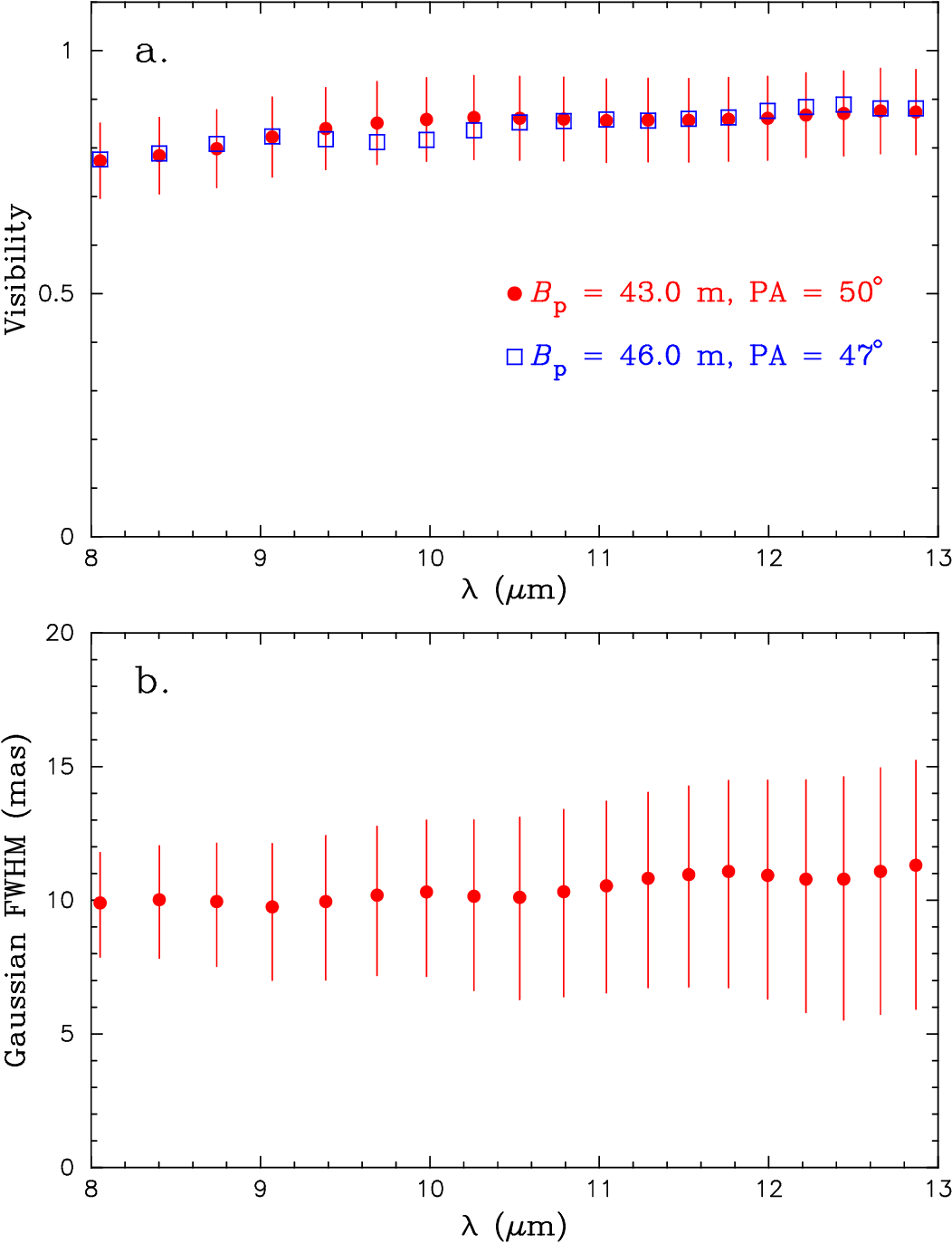}}}
\caption{
VLTI/MIDI observations of \fgsge.
{\bf a:} Observed visibilities of the data set SCI1 (red dots) and
the average of SCI2 and SCI3 (blue dots).
The error bars are only shown for the data set SCI1 for
visual clarity.
{\bf b:} FWHM obtained by a Gaussian fit to the observed visibilities.
}
\label{obsres}
\end{figure}

\begin{figure}[!t]
\resizebox{\hsize}{!}{\rotatebox{-90}{\includegraphics{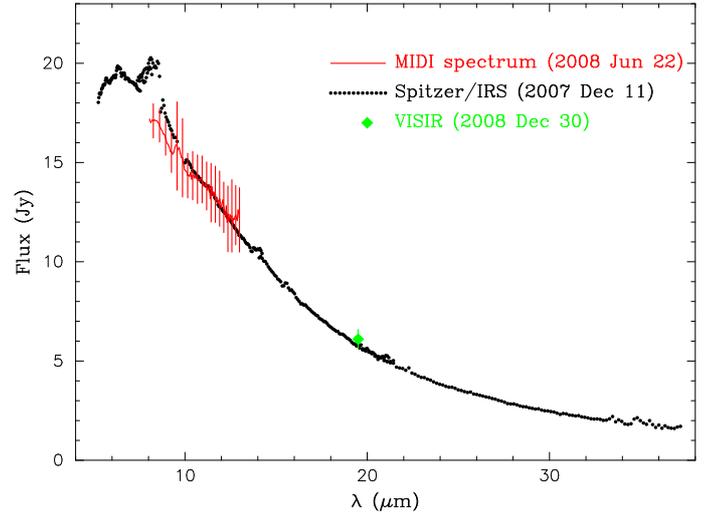}}}
\caption{
Observed mid-infrared spectrum of \fgsge. The solid red line with the error bars shows the MIDI spectrum.
Black dots correspond to the Spitzer/IRS spectrum.
The green diamond is the VISIR Q3-band (19.5~\micron) flux.
}
\label{obsspec}
\end{figure}

\section{Results}
\label{sect_result}

Figure~\ref{obsres}a shows the measured $N$-band visibilities, which 
slightly increase from 0.8 at 8~\micron\ to 0.9 at 13~\micron.
Figure~\ref{obsres}b shows the full width at half maximum (FWHM) obtained
by the Gaussian fitting, which suggests that the observed visibilities
correspond to an approximately constant FWHM of $\sim$10.5~mas. 
The observed visibilities suggest that the dust envelope of \fgsge\ is compact
compared to the other born-again objects' Sakurai's object and V605~Aql. 
Chesneau et al. (\cite{chesneau09}) showed
that the MIDI visibilities of
Sakurai's object obtained at baseline lengths of 42--46~m -- the same as
in our observations of \fgsge\  -- range from 0.05 to 0.2, which is much lower
(thus much more extended) 
than the 0.8--0.9 obtained for \fgsge; although, the distance to two
objects is comparable (\fgsge : 3.0~kpc, Sakurai's object: 3.8~kpc,
Evans et al. \cite{evans20}). 
Clayton et al. (\cite{clayton13}) report that V605~Aql was so large that
it was resolved out in the MIDI observations with UT2 and UT3 (the same
as used for \fgsge) in spite of its distance of 4.6~kpc.
This is primarily because the
effective temperature of \fgsge\ (4500--5500~K,
see Sect.~\ref{sect_model}) is noticeably
lower than those of Sakurai's object (12\,000~K, van Hoof et al.
\cite{vanhoof07}) and V605~Aql (95\,000~K, Clayton et al. \cite{clayton06}). 
In addition, the luminosity of these latter 
objects ($\sim \!\! 10^4$~\LSOL,
Sch\"onberner \cite{schoenberner08}; Chesneau et al. \cite{chesneau09})
is higher than that of
\fgsge\ ($4850^{+4600}_{-2690}$~\LSOL, Sect.~\ref{sect_model}), which can also 
make them appear larger than \fgsge.

Figure~\ref{obsspec} shows the MIDI spectrum (red solid line with the
error bars) and
the 19.5~\micron\ flux measured with VISIR (green filled diamond). 
To complement the MIDI $N$-band spectrum and the VISIR $Q$-band photometric
data, we downloaded the mid-infrared spectrum 
taken with the InfraRed Spectrograph\footnote{
  The IRS was a collaborative venture between Cornell University and Ball
  Aerospace Corporation funded by NASA through the Jet Propulsion Laboratory
  and Ames Research Center.}
(IRS, Houck et al. \cite{houck04}) onboard the Spitzer Space Telescope 
(Werner et al. \cite{werner04})\footnote{Downloaded from
  https://sha.ipac.caltech.edu/applications/\\Spitzer/SHA/}. 
The data were taken on 2007 December 11 (MJD 54446) in the short and
long low mode with spectral resolutions of 64--128, covering from 5 to
40~\micron\ (Program ID: 40061, P.I.: A.~Evans).

The 19.5~\micron\ flux measured with VISIR
is in good agreement with the Spitzer data, which were taken approximately
six months before the MIDI observations. The MIDI spectrum is also
consistent with the Spitzer/IRS spectrum longward of $\sim$10~\micron.
However, the Spitzer/IRS spectrum shows a bump at $\sim$8~\micron, which is not
seen in the MIDI spectrum. 
Evans et al. (\cite{evans15}) reported this feature and the one at
$\sim$6.2~\micron\ as the unidentified infrared 
emission, which is commonly attributed to polycyclic aromatic hydrocarbons
or PAH (e.g., Allamandola et al. \cite{allamandola99}). 
Evans et al. (\cite{evans15}) show that the Spitzer spectrum taken on 2006
October 24 (MJD54032) also exhibit the PAH features at 6.2 and 8~\micron,
which suggests that the PAH emission was stably present over a little
more than a year from October 2006 to December 2007. Therefore, it is not
very likely -- if not entirely excluded -- that the PAH emission
disappeared entirely within six months
between the Spitzer observation in December 2007 and our MIDI observations in
June 2008. The absence of the PAH emission in the MIDI spectrum can be
due to a smaller field of view of 200~mas of MIDI compared to the
slit widths of 3\farcs6--10\farcs7 of Spitzer/IRS, which suggests that
the PAH emission may originate from an extended region. 
  Given that the central star is now H-deficient, the recently rejected inner
  circumstellar matter should also be H-deficient, from which PAH emission
  is not expected. Therefore, the PAH emission likely originates
  from the older,
  outer circumstellar material ejected before the star became H-deficient. 
  This is consistent with the above observations. 
  However, the UV radiation required to give
  rise to the PAH emission cannot penetrate the optically thick dust
  envelope suggested by our modeling presented below.
We return to this point in Sect.~\ref{sect_model}.

\section{Modeling of the dust envelope}
\label{sect_model}

We carried out radiative
transfer modeling with DUSTY (Ivezi\'c et al. \cite{ivezic99})
for the observed $N$-band
visibilities and the spectral energy distribution (SED). 
The circumstellar envelope of \fgsge\ is considered to be complex,
given the R~CrB-like light curve that suggests the formation of clumpy dust
clouds or episodic mass loss. However, the present MIDI data are insufficient
for constraining such complex structures. Therefore, we assumed spherical
shell models to constrain the global properties of the dusty environment
of \fgsge. 

In the present work, we adopted a distance of $3.0^{+1.2}_{-1.0}$~kpc
derived by Chornay et al. (\cite{chornay21})\footnote{The errors in the distance correspond to the 84th and 16th percentile of the distance posterior.} based on the Gaia Early Data Release 3
(Gaia Collaboration \cite{gaia21}). The bolometric flux of the central
star was obtained from the SED obtained
before the onset of the mass ejection in 1992.
Jurcsik \& Montesinos (\cite{jurcsik99}) presented the photometric data from
the $V$ to $N$ band obtained from June to August in 1983, which they fit with a
photospheric model with \TEFF\ = 5500~K and \LOGG\ = 1.5.
We obtained a bolometric flux of $1.72\times 10^{-11}$~W~m$^{-2}$ by 
integrating this SED, which translates into a luminosity of
$4850^{+4600}_{-2690}$~\LSOL\ with the distance of $3.0^{+1.2}_{-1.0}$~kpc.
It should be noted that the large uncertainty in the distance (hence
in the luminosity) does not affect the fit to the SED and visibility,
because the luminosity of the model star increases as $\propto d^2$,
but the flux predicted at the distance of the Earth decreases as
$\propto d^{-2}$, resulting in the same predicted flux. 
Likewise, the inner radius of the dust shell increases as
$\propto \sqrt{L} \propto d$, but its angular size remains unaffected
because it follows $\propto d^{-1}$. 

To obtain the SED observed as contemporaneously as possible with our
MIDI observations, 
we used the visible ($V$, $R$, $I$, and $R_c$ bands) photometric data 
taken on MJD54642 (just three days after our MIDI observations) by
Arkhipova et al. (\cite{arkhipova09}) and 
the infrared ($JHKLM$) photometric data taken on MJD54659.5 (20 days after
the MIDI observations) by Shenavrin et al. (\cite{shenavrin11}). 
As presented above, the MIDI spectrum and VISIR 19.5~\micron\
flux are in agreement with the Spitzer spectrum obtained in December 2007
except for the PAH emission. Therefore, we used these data to cover the
mid-infrared region of the SED. 
  We adopted $E(g-r) = 0.23$ based on the 3D extinction map of
  Green et al. (\cite{green19})\footnote{http://argonaut.skymaps.info}
  for the distance of 3.0~kpc 
  and converted it to $E(B-V) = 0.202$ using the relations given
  in Schlafly et al. (\cite{schlafly11}). This translates into $A_V = 0.63$
  with $R_v = 3.1$ assumed.
  The observed SED was dereddened by applying the
  wavelength dependence of the interstellar extinction from
  Cardelli et al. (\cite{cardelli89}).

The effective temperature of the central star is controversial.
Jeffery \& Sch\"onberner (\cite{jeffery06}) concluded that \fgsge\ had been
maintaining \TEFF $\approx$5500~K since 1980 (as of 1995) and, therefore, 
proposed that this \TEFF\ might be the lowest limit in its redward
evolution. On the other hand, Fadeyev (\cite{fadeyev19}) derived
\TEFF\ = 4445~K based on theoretical modeling of \fgsge's pulsation.
We treated \TEFF\ as a free parameter and calculated models with
\TEFF\ = 4500, 5000, and 5500~K. 
Combined with the luminosity of 4850~\LSOL, these values of \TEFF\ 
correspond to a stellar radius of 115, 93, and 77~\RSOL, respectively. 

Given the C-rich nature of the photosphere and the circumstellar
envelope, we assumed amorphous carbon for the dust grains around \fgsge. 
We adopted a grain size of 0.01~\micron\ as concluded by Arkhipova et al.
(\cite{arkhipova22}) to explain the SED before 2019. 
The density distribution of dust grains was assumed to be $\propto r^{-p}$,
where $p$ is a free parameter, and characterized by  
the dust condensation temperature ($T_c$), which determines the inner
radius of the envelope ($\RIN$), and the optical depth in the radial
direction at 0.55~\micron\ ($\TAUV$).
The outer radius ($\ROUT$) of the circumstellar envelope was estimated
in the following manner. 
Gonzalez et al. (\cite{gonzalez98}) measured an outflow velocity of
$\sim$200~\KMS\ from the \ion{Na}{I} D lines.
Because the mass ejection started in August 1992, the material should have
traveled $1.0\times10^{16}$~cm (670~AU) at the time of our MIDI observations 
in June 2008.
On the other hand, DUSTY models show that the inner radius of the envelope
ranges from 25 to 40~\RSTAR\ for \TEFF\ = 4500 to 5500~K for a 
condensation temperature of $\sim$1000~K. 
Using the above values of \RSTAR, the inner radius is estimated to be
$(1-3) \times 10^{14}$~cm.
Therefore, the maximum distance that the
ejected material traveled ($1.0 \times 10^{16}$~cm) corresponds to
33--100~\RIN. We adopted a fixed value of 70~\RIN.

We computed models with \TEFF\ = 4500, 5000, and 5500~K, 
$p$ = 2.0 ... 4.0 ($\Delta p$ = 0.5),
$T_c$ = 800 ... 1600~K ($\Delta T_c = 200$~K), and 
\TAUV\ = 5 ... 30 ($\Delta \TAUV = 5$). 
However, 
we found no model that can reasonably reproduce the observed SED and $N$-band 
visibilities simultaneously. For example, 
the models with the density profile $\propto r^{-2}$, which corresponds to
stationary mass loss with a constant outflow velocity, can fit 
the observed SED very well,
but they systematically predict the $N$-band visibility
to be much lower ($\la$0.5) than the observed data, because the dust
envelope appears to be much more extended than the observed data suggest. 
The models with steeper density profiles can explain the observed $N$-band
visibilities. However, these models predict the flux longward of
$\sim$10~\micron\ to be too low compared to the observed data,
because there is
too little cold material in the outer region of the envelope.

\begin{figure}[!t]
\resizebox{\hsize}{!}{\rotatebox{0}{\includegraphics{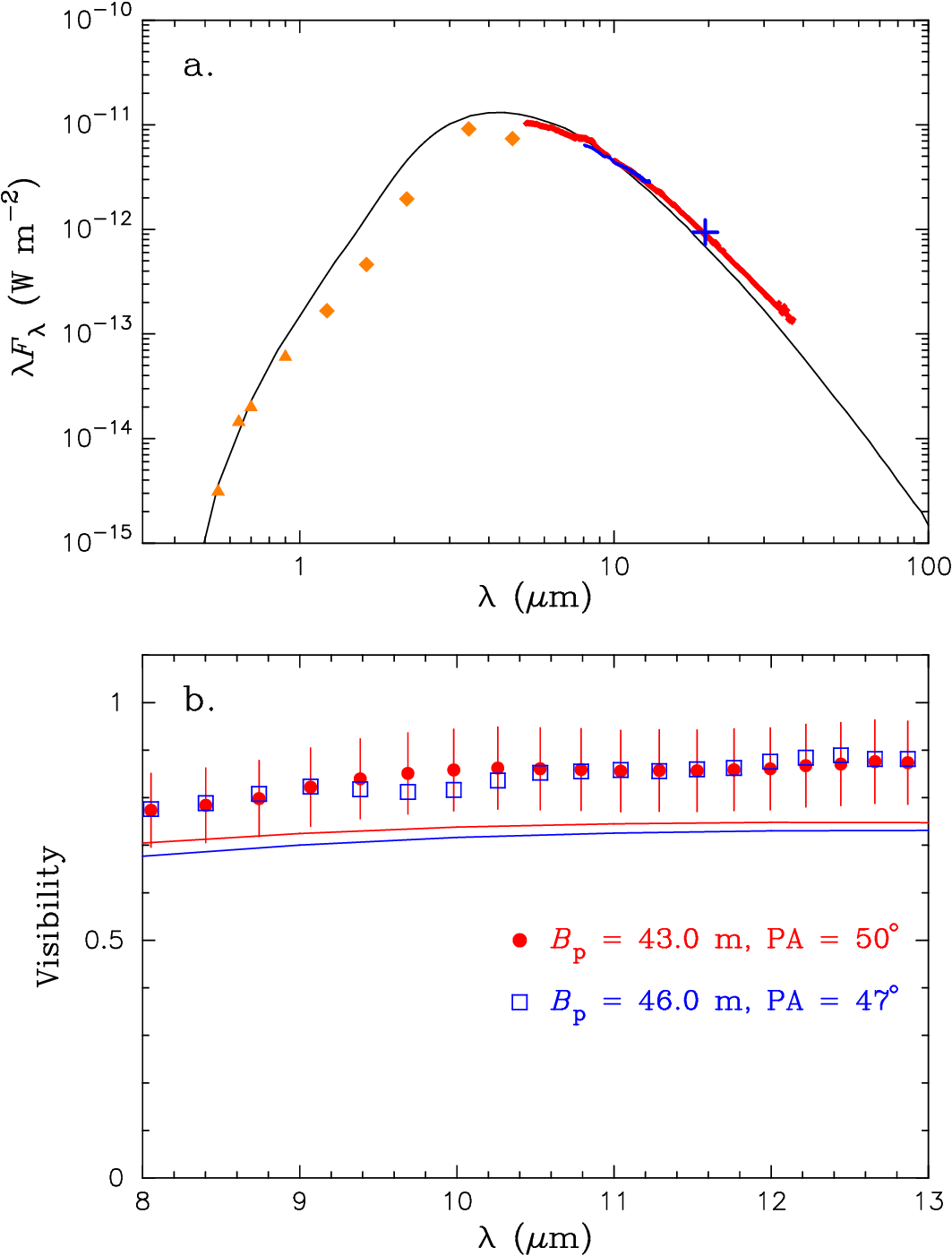}}}
\caption{
Comparison of a spherical shell model with 
the observed SED and $N$-band visibilities of \fgsge.
The model is characterized with \TIN\ = 1100~K, a piecewise power-law
density profile with $r^{-3.5}$ ($\RIN \le r \le \RTRANS$ with \RIN\ = 
33~\RSTAR\ and \RTRANS\ = 5~\RIN) and $r^{-2}$ ($r > \RTRANS$),
\TAUV\ = 8, and \TEFF\ = 5000~K. 
{\bf a:} Observed SED.
Orange filled triangles correspond to visible photometric data from Arkhipova et al.
(\cite{arkhipova09}).
Orange filled diamonds are the infrared photometric data from Shenavrin et al.
(\cite{shenavrin11}).
The photometric data were dereddened with $A_V$ = 0.63. 
The thick solid red line is the Spitzer/IRS spectrum.
The thin solid blue line corresponds to the MIDI spectrum.
The blue cross corresponds to VISIR photometry.
The DUSTY model is shown with the solid black line.
{\bf b:} Observed visibilities are shown in the same manner as in
Fig.~\ref{obsres}a. The solid red and blue lines represent the
model visibilities predicted for the 43.0 and 46.0~m baselines, respectively. 
}
\label{dusty_model}
\end{figure}

It is possible that the density profile is steeper than the $r^{-2}$
in the inner region of the dust envelope before the velocity reaches its
terminal velocity. We attempted to explained the observed data with
models in which the density profile falls as $r^{-p}$ from the inner
radius \RIN\ to a certain radius \RTRANS\ and then changes to $r^{-2}$.
Because the above model grid shows that \TAUV\ should be below $\sim$10
to explain the visible part of the observed SED, 
we calculated models with the following parameter ranges:
\TEFF\ (K) = 4500, 5000, and 5000; 
\TAUV\ = 4.0 ... 10.0 ($\Delta \TAUV\ = 2.0$); \TIN\ (K) = 800 ... 1600
($\Delta \TIN$ = 100~K); $p$ = 2.5 ... 4.5 ($\Delta p = 0.5$); and 
\RTRANS /\RIN\ = 2.5, 5, 10, 15, and 20.
Figure~\ref{dusty_model} shows one of the best-fit models,
which
reproduces the observed SED and $N$-band visibilities fairly well. 
The model is characterized by \TEFF\ = 5000~K, \TAUV\ = 8, \TIN\ = 1100~K
(\RIN\ = 33~\RSTAR), $p$ = 3.5, and \RTRANS\ = 5~\RIN. 
The model still predicts the flux to be somewhat lower than the observed
data longward of 20~\micron, and the model visibilities are slightly lower
than the MIDI data. Decreasing $p$ and/or \RTRANS\ makes the fit to the
SED better, but such models predict the $N$-band visibilities to be even
lower, because their density profiles become similar to $r^{-2}$.
Our model grid constrains the parameter ranges
as follows: \TAUV\ = $8\pm 2$, \TIN\ = $1100 \pm 100$~K, $p = 3.5 \pm 0.5$, and
\RTRANS\ = $10 \pm 5$. 
We cannot constrain \TEFF\ of the central star within the adopted range
between 4500 and 5500~K, because it is heavily obscured. 

Recently, Arkhipova et al. (\cite{arkhipova22}) modeled the SED observed from 0.4 to 5.0~\micron\ in 2019 with a
density profile of $\propto r^{-2}$ and obtained
an optical depth of 4.5 at 0.55~\micron, which is significantly lower than
the $8 \pm 2$ derived in the present work. 
This is because they modeled the photometric data taken in 2019, when the
dust clouds temporarily cleared. 
  The central star became visible
  with $V \approx 13$, which is $\sim$4 magnitude fainter than the 
  $V \approx 9$ before the start of the dust formation in 1992.
  With $\TAUV = 0.921 \times \Delta V$ (if scattered light and
  dust thermal emission are ignored), \TAUV\ is estimated to be 3.7,
  which is in broad agreement with 
  4.5 as derived from their modeling. 
As mentioned above, the models with the density profile $r^{-2}$ can
explain the observed SED, but not the $N$-band visibilities. 
Given that Arkhipova et al. (\cite{arkhipova22}) only modeled the SED
taken at a different epoch (May 2019) from ours, 
we cannot draw the conclusion that the difference in the density profile
reflects the lack of spatially resolved data in the modeling 
or a temporal variation in the density profile. 

The remaining discrepancy between the best-fit models and the observed data
is likely due to the complex, clumpy nature of the circumstellar environment
of \fgsge\ as inferred from the irregular, R~CrB-like declines in the
light curve. Furthermore, a disk-like or ring-like 
structure may also be present,
as observed in the other two born-again objects' Sakurai's object and V605~Aql
(Chesneau et al. \cite{chesneau09}; Hinkle et al. \cite{hinkle14},
\cite{hinkle20}; Clayton et al. \cite{clayton13}). 
As described in Sect.~\ref{sect_obs}, 
if the PAH emission indeed originates from the outer region of the
circumstellar envelope, the UV radiation should escape the optically thick
envelope. 
The clumpy and/or disk-like structures can provide a natural
explanation for the UV radiation to escape and excite the carrier
of the PAH emission even if it is located farther away from the star.

\section{Discussion and concluding remarks}
\label{sect_discuss}

The dust mass obtained from the model shown in Fig.~\ref{dusty_model} is
$7.4\times10^{-7}$~\MSOL, adopting a bulk density of 2.25~g~cm$^{-3}$ for
amorphous carbon (Gilman \cite{gilman74}).
If we assume a gas-to-dust ratio of 200, the total 
envelope mass is $1.5\times10^{-4}$~\MSOL.
Given that the mass ejection started in
1992, the average mass-loss rate at the time of the MIDI observations in
2008 is $9.3\times10^{-6}$~\MSOLPERYR. 
The dust mass of \fgsge\ seems to be lower than 
$6 \times 10^{-5}$~\MSOL\ as found in Sakurai's object (Chesneau et al.
\cite{chesneau09}) and $\sim \! 2\times10^{-3}$~\MSOL\ as derived for
V605~Aql (Clayton et al. \cite{clayton13}).
The difference in the ejected mass may be related to the
difference
  in the nature between \fgsge\ and the other two final-flash
  objects -- Sakurai's object and V605~Aql are VLTP objects, while
  \fgsge\ is likely to be an LTP object. 
However, follow-up observations of \fgsge\ would be
necessary to study the long-term variation of the mass of the ejected dust.

As the light curve published by Arkhipova et al. (\cite{arkhipova22}) shows,
\fgsge\ is still experiencing the continuous ejection of dusty material.
If we adopt the outflow velocity of 200~\KMS\ as inferred from the
Na I D lines, the material at the outer radius of the dust envelope
in 2008 should have traveled 560~au ($\sim$1200~\RSTAR) by now, which
translates into an angular displacement of 190~mas at the distance of
3.0~kpc. While the farthest (thus the coldest) material at the outer
boundary may not be detectable in the 10~\micron\ region, temporal variations
in the circumstellar envelope can be expected. 
Furthermore, 
given the presence of disk-like structures in Sakurai's object and V605~Aql,
it is of great interest to study whether \fgsge\ has a disk or bipolar
structures.
To better constrain the structure and evolution of the circumstellar
environment of \fgsge, interferometric measurements at more baselines
and at multiple epochs are indispensable, which can be carried out with
the VLTI/MATISSE instrument (Lopez et al. \cite{lopez22}).

\begin{acknowledgement}
We thank the VLTI team at ESO and the MIDI team for carrying out the
observations and making the data reduction software publicly available.  
K.O. acknowledges the support of the Agencia Nacional de 
Investigaci\'on Cient\'ifica y Desarrollo (ANID) through
the FONDECYT Regular grant 1210652.
This research made use of the \mbox{SIMBAD} database, 
operated at the CDS, Strasbourg, France.

\end{acknowledgement}

\end{document}